# Inventions on presenting textual items in Graphical User Interface
## A TRIZ based analysis

**Umakant Mishra**

Bangalore, India

http://umakantm.blogspot.in

**Contents**

## 1. Introduction

The development of graphical user interface has made the job of a computer user easy and simple. The conventional complex commands are replaced with menus, buttons and various other tools. Although in most cases the textual descriptions are replaced by graphical icons, the textual items are not completely removed. The textual items are inevitably used in window titles, message boxes, help items, menu items and popup items.

Textual items are necessary for communicating messages that are beyond the limitation of graphical messages. But it is necessary to harness the textual items on the graphical interface in such a way that they complement each other to produce the best effect. One may keep the following considerations in mind while applying textual items in graphical user interface.

- ☞ The look and feel of the text should match with the look and feel of the GUI.

- ☞ The textual presentation should supplement the graphics and vice versa.

- ☞ The text should be precise, readable and meaningful.

- ☞ The text need not be shown for messages already conveyed by graphic components.

- ☞ The text can be shown only when necessary and not always, and so on.

Although the developer keeps all the above basic principles in mind, it is difficult to achieve the optimal result because of several contradictions as below:

- ☞ The textual items should be ***elaborate*** to express the intended meaning and at the same time ***precise*** to save the screen space.

- ☞ The size of the text should be ***small enough*** to save screen space and ***large enough*** to ensure readability.

- ☞ The textual items (like tool tips) ***should be displayed*** to explain the graphical components but they ***should not obstacle*** other items on the screen.

- ☞ The list-boxes should contain ***maximum textual items*** within ***minimum screen space*** and provide easy searching.

It is necessary to resolve the above (and other) contradictions in order to display textual items effectively on a Graphical User Interface. Some methods of solving contradictions may be as below.

- ☞ Display textual items (like tool tips) on a specific isolated area of the screen (Principle-2: Taking out),

- ☞ Display text in motion or scrolling mode (Principle-15: Dynamize).

- ☞ Use dynamically changing (contextual) text instead of static or fixed text (Principle-15: Dynamize).

- ☞ Display text in different directions or dimensions, horizontal or vertical, two dimensions or three dimensions (Principle-17: Another Dimension).

The following are some inventions from US Patent database, which shows various innovative methods to improve the display of textual items on a graphical user interface.

# 2. Inventions on displaying Textual items on a graphic user interface

## 2.1 Dynamically updating the title of a window (5784057)

**Background problem**

In some graphical interfaces, the title of the window contains the object(s) displayed in the window. (The objects can be container objects, or data objects or device objects.) In case of conventional static titles, when objects in the window are changed, the title of the window does not match with the contents of the window. This creates confusion to the user as the title denotes something different from the contents of the container.

**Solution provided by the invention**

Patent 5784057 (invented by Alimpich et al., assigned to IBM, Jul 98) discloses a method of dynamically updating the title of a window to correspond with changes in the objects inside the window. The revised title of the window is saved for future use of the window. Besides, the window is automatically closed when all the objects are deleted.

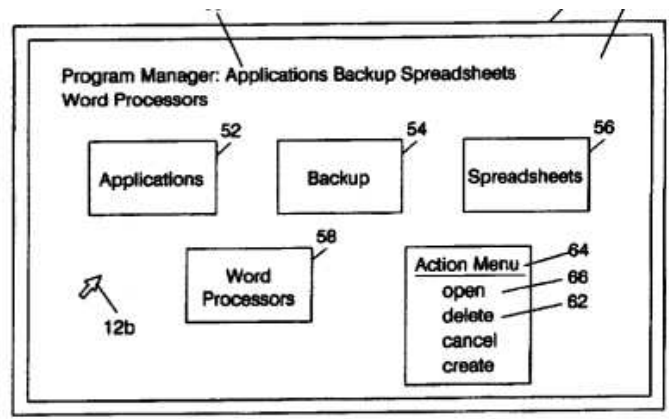

**TRIZ based analysis**

The title of the window is comprised of all the parent objects in the window (Principle-5: Merging).

The invention dynamically updates the title of the window when the objects in the window are changed (Principle-15: Dynamize).

## 2.2 Graphical user interface system and methods for improved user feedback (US Patent 6005570)

### Background problem

Typically a GUI displays several graphic elements like icons and buttons, which are internally connected to specific functions. It is necessary to indicate the function of the graphic elements to the user. But the space on the icon or button is not sufficient to display the name of the function.

A smart icon provides a solution to this problem by displaying a balloon help describing its function when a user moves the cursor on to the smart icon. But this has a major drawback as it obscures the screen by popping up a message at a screen location that is proximate to the cursor location. Secondly, the balloons are resource intensive to frequently repaint the user image.

### Solution provided by the invention

Patent 6005570 (invented by Gayraud et al., assigned by Inprise Corporation, issued Dec 1999) discloses an improved method of displaying icon descriptors without obscuring the active screen. According to the present invention, the "hints" for the icons are displayed along one side of the screen (preferably at the bottom of the screen), which does not block any valuable portion of the screen.

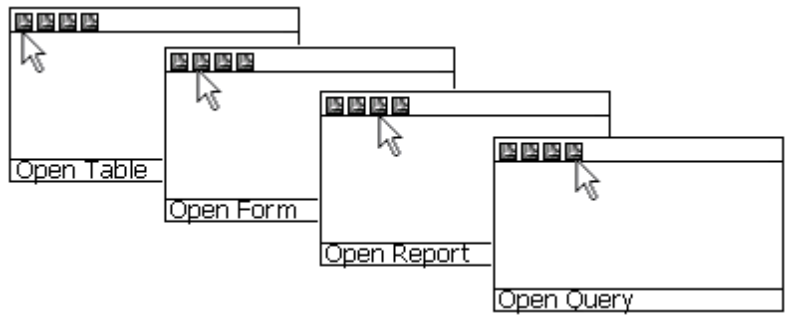

When the cursor enters into the boundary of an object, a corresponding descriptor or "hint" is displayed at a status frame of the window, and when the cursor moves out of the boundary of that object the descriptor is cleared.

### TRIZ based analysis

The invention isolates the descriptor display area from the main working area and displays the descriptions in a static frame at the bottom of the screen, which do not obscure the valuable screen items **(Principle-2: Taking out)**.

The invention displays the descriptor when the cursor moves on to the icon and clears the descriptor when the cursor moves out of the icon **(Principle-15: Dynamize)**.

## 2.3 Next/ current/ last ticker graphical presentation method (6182098)

**Background problem**

In a ticker tape the information scrolls horizontally across a computer display screen. When the user sees a headline of interest, he moves a pointer to the headline and clicks to retrieve a news story associated with the headline. The advantages of ticker display are (i) they are attractive; (ii) they can show a lot of information in a small horizontal space. But they also have disadvantages (i) reading moving information is distracting; (ii) increases eye stress and fatigue.

**Solution provided by the invention**

Patent 6182098 (invented by Selker, assigned by IBM, issued Jan 2001) discloses a method of displaying headlines through a solid polygon. According to the invention the polygon rotates at predetermined intervals, so that the successor headline moves into a dominant position, the previous headline moves to a predecessor position and a new successor headline is displayed. The predecessor and successor headlines are foreshortened to enhance the 3D look of the polygon.

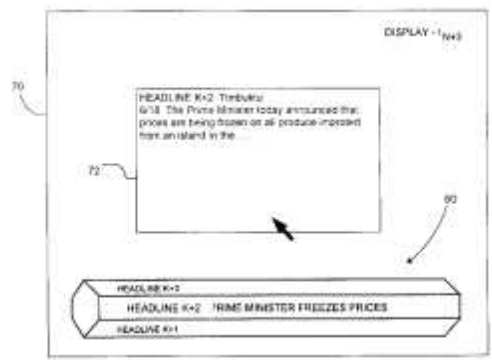

This method is easy to read because the eye does not have to follow the moving text. The user can also click on headlines to do interactive operations.

**TRIZ based analysis**

The invention proposes an upward scrolling of information compared to the leftward scrolling of ticker tapes (Principle-17: Another dimension).

The information is displayed on a three dimensional polygon compared to the conventional linear arrangement (Principle-17: Another dimension).

The polygon rotates at predefined intervals unlike continuous running of tickers (Principle-19: Periodic action).

## 2.4 Portable information terminal and information scrolling method for use therewith (6201524)

### Background problem
With the advancement of telecommunication systems, there is a growth in use of portable telephones and pagers for character based communication. As the screen of a pager is very small and there are limited number of control keys, the information scroll is generally permitted in the downward direction only. However, users feel constrained when they can scroll information in a single direction alone. The information should preferably be scrolled up and down to provide convenience.

### Solution provided by the invention
Patent 6201524 (invented by Aizawa, assigned by Sony Corporation, issued Mar 2001) solves the above problem by providing two scroll modes, viz., a line scroll mode or a screen scroll mode, without providing additional control keys.

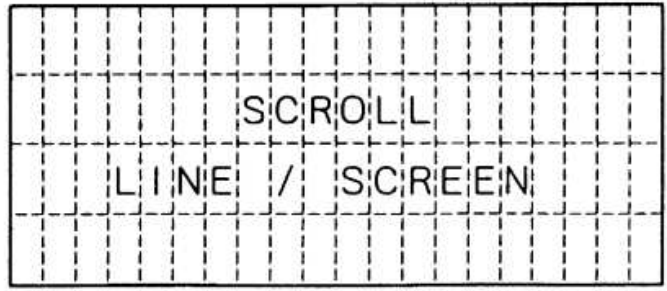

### TRIZ based analysis
The invention uses two types of scrolling, viz., line scrolling and screen scrolling, compared to the one directional scrolling of the prior art (Principle-17: Another dimension).

## 2.5 Method, system and computer program product for rotating through a sequence of display states in a multi-field text string class in a graphical user interface (6388686)

### Background problem
A list box or dropdown box allows a user to choose an item among a group of display items. However, a dropdown box obscures the display of important information. Besides, the user may not know which item in the dropdown list will produce the desired effect requiring the user to successively try a number of display items until the desired item is identified. For example, one may have to try several times to obtain the desired clipart image if he is unable to identify the correct image by name.

**Solution provided by the invention**

Patent 6388686 (invented by Hetherington et al., assigned to IBM, issued May 2002) provides a rotate control for selecting or altering a user interface display item. According to the invention the rotate control displays a group of display items. When the user actuates the rotate control, it displays item by item in a sequence. An indicator provides a visual cue to the user of which display item is currently selected for display.

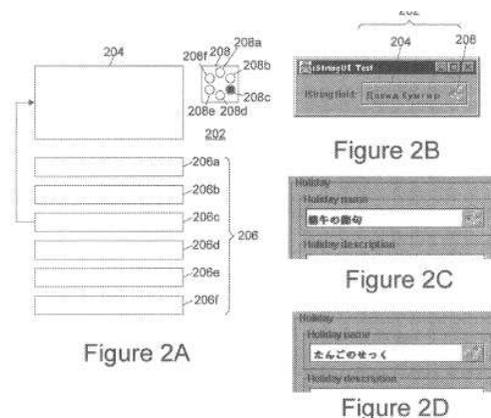

Figure 2A, Figure 2B, Figure 2C, Figure 2D

This mechanism is useful for selecting text strings from a small group of display items. The rotate control provides a faster switching between views than the conventional dropdown boxes. Besides the rotate control provides a fixed view unlike flyover popup box displays.

**TRIZ based analysis**

The rotate control displays items in a rotary view compared to the dropdown list that displays items in a linear view (Principle-14: Curvature).

The rotate control provides a fixed view without obscuring the screen information (Principle-39: Calm).

**2.6 Ticker display on an active desktop (6421694)**

**Background problem**

With the growing need of information requirement, it is often found that the limited screen space is not enough to display the information required. Although a desktop metaphor uses icons on the desktop to click and get various types of information, still the method has limitations of the screen space.

**Solution provided by the invention**

Nawaz et al. provided a method for displaying tickers on the window (Patent-6421694, Assigned to Microsoft Corporation, Jul 2002). According to the invention, the data is displayed continuously on ticker display pane of the client computer. The data items are displayed in a continuous sequence and may be provided from Internet servers, Intranet servers or other sources.

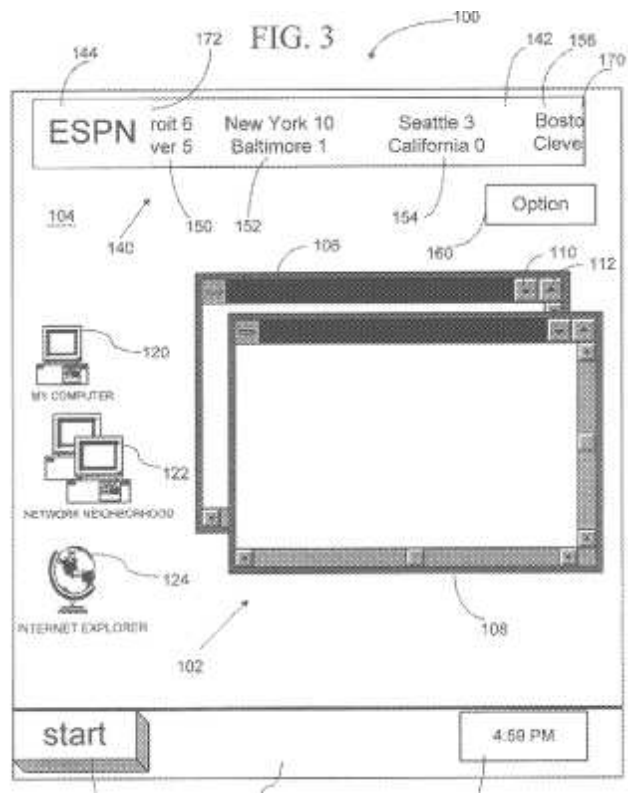

**TRIZ based analysis**

The invention uses tickers on a ticker display pane for dynamically displaying data items (Principle-15: Dynamize).

The ticker pane continuously displays data items, thereby adding a new dimension that displays more data in a limited screen space (Principle-17: Another dimension).

## 3. Summary

It is not only the font size, style and color that should be taken care to match with the GUI but a lot of other issues are important in incorporating textual items effectively in a Graphical User Interface, such as place of display, timing of display, duration of display, conditions of display and many more.

The difficulties of displaying textual items in a Graphical User Interface are eliminated by various methods illustrated in the above inventions. In some cases the text is made flexible to change its contents, in other cases it is made dynamic to scroll and in other cases the text may appear and disappear as per the need. The future days will find more and more inventions on this issue for better and more effective presentation of textual items in a Graphical User Interface.

# Reference:

1. US Patent 5784057, "Dynamically modifying a graphical user interface window title", invented by Alimpich et al., assigned to IBM, Jul 98.

2. US Patent 6005570, "Graphical user interface system and methods for improved user feedback", invented by Gayraud et al., assigned by Inprise Corporation, issued Dec 1999.

3. US Patent 6182098, "Next/current/last ticker graphical presentation method", invented by Selker, assigned by IBM, issued Jan 2001.

4. US Patent 6201524, "Portable information terminal and information scrolling method for use therewith", invented by Aizawa, assigned by Sony Corporation, issued Mar 2001.

5. US Patent 6388686, "Method, system and computer program product for rotating through a sequence of display states in a multi-field text string class in a graphical user interface", invented by Hetherington et al., assigned to IBM, issued May 2002.

6. US Patent 6421694, "System and method for displaying data items in a ticker display pane on a client computer", invented by Nawaz et al., Assigned to Microsoft Corporation, Jul 2002.

7. US Patent and Trademark Office (USPTO) site, http://www.uspto.gov/